\documentstyle[epsfig]{aipproc}

\def\simlt{\lower.5ex\hbox{$\; \buildrel < \over \sim \;$}}

\begin{document}

\title{Theory of Protostellar Accretion}
 
\author{Eve C. Ostriker}
\address{Department of Astronomy\\
University of Maryland, College Park, MD 20742-2421\\
ostriker@astro.umd.edu}

\maketitle

\begin{abstract}
This paper reviews current theoretical work on the various stages of 
accretion in protostars, and the relationship of these ideal
stages to the spectral classes of observed protostellar systems.  I
discuss scaling relationships that have been obtained for models of 
pre-stellar cores as they evolve by ambipolar diffusion toward a central 
singularity, and expectations for the dynamical evolution as the cores 
collapse radially to form a rotating disk.  I summarize work that suggests 
accretion in T Tauri systems may be limited by ionization to 
disk surface layers, and implications for the variation of $\dot M_D$ with
radius.  Finally, I describe models for the asymptotic 
structure of free magnetocentrifugal winds, and show that the constant-density 
surfaces in
these winds may be strongly collimated even when the streamlines are not.
\end{abstract}

\section*{Introduction}

The subject of this review is accretion in protostellar systems, which
is properly a hot topic in a figurative but not a literal sense.  In
the other accreting systems discussed at this meeting, the observed
high-energy emission often dominates the longer-wavelength flux, and 
much of the current theoretical effort is devoted to developing
dynamical and radiative models of {\it hot} accretion flows that are capable
of producing the observed spectral distributions.
For low-mass protostars and T Tauri stars, on the other hand, 
the observed X-ray emission of $10^{29}-10^{31}$
erg $s^{-1}$ \cite{cas95,koy96,mon93,neu95} is much enhanced relative
to that of main-sequence stars, but still constitutes just a small
fraction ($10^{-4} -10^{-3}$) of the sources' bolometric luminosities;
it can be understood as thermal emission from $T\sim 10^6-10^7$K
plasma in the stellar magnetospheres and inner parts of the jets of
these systems \cite{fei91,fei97}.  The bulk of the emission from young,
low-mass stellar objects (YSOs) originates instead in a cooler accretion flow,
where gas temperatures range from $\sim 10^4$ K (in the
boundary layer/hot spot region), to a few $\times 10^3$K (in the
inner, centrifugally-supported, accretion disk) to $\sim 10$K (in the
outer, radial parts of the flow).  As will be described below, many of
the current issues facing theorists interested in protostars have a
quite different character from the issues under investigation in
higher-energy systems.

In this paper, I will begin by briefly summarizing the standard
observational classification of low-mass YSOs, which assigns sources
to one of four broad groups based on their spectral energy
distributions.  As Nuria Calvet describes in detail in her
observational review in this volume, other properties of YSOs -- such
as the relative masses of their disks {\it vs.} envelopes, their
accretion rates, their outflow momentum flux, etc. -- also correlate
with their spectral properties.  The spectral classification is
generally interpreted as an evolutionary sequence corresponding to
successive stages of protostellar accretion
\cite{shu87,ada87,lad95,shu95a}; this paper outlines current
understanding of the underlying theory.  Since space constraints make
a comprehensive review impossible (see instead the volume \cite{pp3}
and its successors), I will focus on some of the ways in which
protostellar accretion is {\it different} from the other sorts of
accretion discussed at this meeting.  I will highlight some recent
results in the theory of the early (i.e. largest-scale) stages of
protostellar accretion, in which the flow is primarily radial.  I will
also describe recent work suggesting that even Shakura-Sunyaev-type
``$\alpha$''-models of the late, T-Tauri phase may be quite different
from the standard (e.g. \cite{pri81}) $\dot M_D=constant$ paradigm.  In the
last section, I will describe models of free, wide-angle YSO
winds that are different from the confined, narrow jets
found in high-energy systems (cf. the review of Blandford in this
volume).

\section*{Observational Classes of Young Stars}

The spectral classification of YSOs introduced by Lada \cite{lad87} and 
expanded by Andre, Ward-Thompson, \& Barsony \cite{and93} divides them into
classes 0-III, as follows: 

{\bf Class 0}:  These are embedded objects whose spectral energy distributions
(SEDs) peak in the sub-millimeter and show little or no IR, and no optical.  
They are believed to be the earliest protostars, since the masses of their
envelopes $0.2-2 M_\odot$ exceed the masses of their disks + protostars 
\cite{and93,and94,bon96}.  From the relative numbers of these sources
compared to other YSO, ages are probably a few $\times 10^4$ years.

{\bf Class I}:  These are embedded objects which are strong far-IR emitters,
with a peak in the SED at $10-100\mu$.  They are believed to correspond to
late protostars or ``pre-T-Tauri'' stars since their envelope masses are
typically $0.1-0.3M_\odot$ or less \cite{and94}. From the relative number of 
these sources, 
ages are estimated at a few $\times 10^5$ years (e.g.\cite{wil89}).

{\bf Class II}:  These are optically-revealed objects whose spectral energy
distributions consist of a stellar photosphere plus significant ``excess'' IR 
and UV.  These are interpreted as systems containing a star, an accreting
disk, a boundary layer/hot spot, and a remnant envelope.  Disk masses derived
from mm continuum during this stage are 0.01-0.1 $M_\odot$ \cite{bec90}
Ages of classical T Tauri stars (CTTS) 
(nearly coincident with this class) are $1-4\times 10^6$ yrs
\cite{coh79}.

{\bf Class III}:  These sources present stellar blackbodies, with no sign
of active accretion; this class is essentially the same as the weak T Tauri 
stars (WTTS).

\section*{Stages of Protostellar Accretion}

\subsection*{Ambipolar Diffusion and Density Restructuring}

Molecular clouds in our Galaxy contain hundreds of thermal Jeans
masses ($M_J=c_s^3(\pi/G)^{3/2} \rho^{-1/2}$); the fact that the
observed star formation rate is orders of magnitude below the ratio of
the gas mass to the gravitational collapse time is used to argue that
molecular clouds are primarily supported by magnetic pressure and
tension rather than thermal pressure (e.g. \cite{shu87,mck93,ost97a}).
Under these circumstances, dynamical collapse to form a star occurs
only when the ratio of mass to magnetic flux exceeds a certain
critical value \cite{mou76,tom88}; for the central flux tube the
corresponding ratio of surface density to magnetic field is $\Sigma/B=
1/(2\pi \sqrt{G})$.  A given column density for a cloud implies a
certain critical magnetic field strength $B_{crit}$, and vice versa.
Using a field strength of $20\mu G$, and a size scale of $\sim0.1$pc,
the critical mass (for a spherical cloud) corresponds to $\sim
M_\odot$, comparable to the minimum masses of dense cores where low
mass stars are observed to be forming.

As Mestel and Spitzer pointed out in the 1950s, the first stage in the
protostellar accretion process under present Galactic conditions is
expected to be the inward diffusion of neutrals relative to ions and
the magnetic field, in order to create a supercritical core within a
subcritical cloud.  Thus the first accretion timescale is the
ambipolar diffusion timescale $t_{AD}=L/v_{drift}$, which can be
estimated by equating the drag force on ions from inflowing neutrals
with the magnetic force on ions; the result is \cite{mou89}
\begin{equation}
t_{AD}\approx 2\times 10^6 {\rm yrs} \left({n_i/ n_{H_2}\over 10^{-7}}\right)
\left({B_{crit}\over B} \right)^2.
\end{equation}

The ambipolar-diffusion/density restructuring stage has been studied by
Nakano \cite{nak79} and Lizano and Shu \cite{liz89} in the quasistatic limit
(applicable to early phases), and by Mouschovias and his collaborators
\cite{fie93,cio94,bas94} with a full dynamical treatment.  For the 
latter numerical calculations, evolution is typically followed until the
central density reaches $10^9$ cm$^{-3}$, 1-2 orders of magnitude below the
limit at which an opaque ``protostar'' would form.  The same 
phase of evolution of initially subcritical cores has also recently been 
studied semi-analytically by modeling magnetic stresses in an angle-averaged 
sense \cite{saf97}.  In complementary studies, Tomisaka \cite{tom96} 
has performed ideal-MHD (i.e. no ambipolar diffusion) simulations of
 the evolution 
of magnetized clouds which are initially supercritical, beginning from a 
cylindrically-stratified equilibrium.  While the results of all these authors
differ in detail depending on their assumptions and parameter regime, the
similarities in the results are more striking;  I will concentrate on the
latter.  In both the initially-subcritical and the initially-supercritical
simulations, the model clouds evolve towards flattened (disk-like), infalling
structures in which the surface density $\Sigma\propto 1/R$, the magnetic 
field strength $B\propto 1/R$, and the midplane density 
$\rho_c\propto 1/R^{2}$, where $R$ is the cylindrical radial coordinate.  The
peak inflow speed is 2-3 times the sound speed (which for $T=10$K is 
$c_s=0.2$ km s$^{-1}$).  

Schematically, these common properties
can be understood from dimensional analysis as follows:  First, marginally
supercritical disks obey $B=2\pi G^{1/2} \Sigma$ locally, so that once a
region of the cloud becomes supercritical it will maintain $B\propto \Sigma$
because the ambipolar diffusion time is long compared to the dynamical time.
Second, in local vertical equilibrium, an isothermal disk has central density
$\rho_c=\pi G \Sigma^2/(2c_s^2)$ \cite{spi78}; hence we can expect 
$B\propto\Sigma\propto \rho_c^{1/2}$ for the dynamically-evolving part of the
flow.  Finally, near-balance of the inward force of gravity 
($\sim GM/R^2\sim G\Sigma$)
against the large-scale gradients of magnetic pressure and tension 
($\sim B^2/(\rho_c R)\rightarrow G \Sigma^2/(\rho_c R)\rightarrow c_s^2/R$) 
and thermal pressure ($\sim \rho_c c_s^2/(\rho_c R)\rightarrow c_s^2/R$) 
would imply $\Sigma \sim c_s^2/(G R)$, and hence 
$B\propto 1/R$ and $\rho_c\propto 1/R^{2}$.  Ideas along these lines have
been developed with more rigor by Basu \cite{bas97} (see also
this volume), who constructs an approximate self-similar dynamical solution 
that matches well to his numerical solutions for this stage of accretion.

\subsection*{Dynamical Collapse and Centrifugal Disk Formation}

What comes next?  How does dynamical collapse proceed in the
centrally-condensed structures, so as to appear as the observed Class
0 and Class I objects?  At this point, there is not the same
convergence of theoretical models for the dynamical collapse stage (in
which speeds approach free-fall in the inner parts) as there is for
previous restructuring stage (which involves speeds no greater than a
few km s$^{-1}$).  Because the result of the previous stage yields a
density profile $\rho\propto r^{-2}$, though, we may anticipate close
resemblance to the family of (unmagnetized) isothermal-sphere
self-similar collapse models whose best-known representatives are the
Larson \cite{lar69}/Penston \cite{pen69} and Shu \cite{shu77} solutions.

To gain insight into the dynamical collapse process, Galli \& Shu \cite{gal93}
calculated solutions for collapsing clouds as perturbations with nonzero
magnetic field strength about the inside-out singular isothermal sphere 
collapse.  They found that, even with an initially uniform magnetic field 
and spherical density distribution, magnetic pinching forces lead to the
formation of a growing equatorial ``pseudodisk.'' A small-amplitude 
magnetic field, however, does not alter the accretion rate;  this remains 
equal to the value $0.975 c_s^3/G$ of \cite{shu77}.  In an opposite limit,
Li \& Shu \cite{li97} demonstrate that an initially static disk with initial 
magnetic and surface density profiles 
$B\propto \Sigma\propto 1/R$ (i.e. the asymptotic scalings obtained from the
evolution in the ``restructuring'' stage) undergoes an inside-out collapse.
The accretion rate is given by $\dot M\approx (1+H_0)c_s^3/G,$ where $1+H_0$
is the density overfactor allowed by magnetic support in the initial 
equilibrium compared to an unmagnetized state ($H_0=0$).  Although the 
calculation was performed in the $H_0>>1$ thin-disk limit, more realistic 
cases where $H_0=1-10$ may be expected to show similar behavior for the 
dynamical collapse that follows the formation of a  central density 
singularity.  

In the very innermost regions, the solution must approach free-fall,
and the asymptotic collapse solution for a perfectly cold,
unmagnetized disk with initial surface density $\Sigma_i \propto
R^{-1}$ is informative.  For disk with initial profile
$\Sigma_i(R)=M_{i}(R)/(2\pi R^2)\propto R^{-1}$ where $M_i\propto R$,
the resulting asymptotic scalings (for large time $t$ and small radius
$R$) are $\Sigma \propto t^{-1/2}R^{-1/2}$, $v\propto
t^{1/2}R^{-1/2}$, and $\dot{M} = const.=M_{i}(R)/t_{collapse}(R)= 2^3
(\pi G)^{1/2} (\Sigma_i (R) R)^{3/2}$.  Using the surface density
profile $\Sigma\approx c_s^2/(GR)$ of \cite{bas94}, one obtains
$\dot{M}=14c_s^3/G$, or about $2\times 10^{-5} M_\odot/ {\rm yr}$ for
typical conditions.  The associated collapse time $\sim 10^5{\rm yr}$
is comparable to the observationally-estimated ages of embedded
sources, giving some confidence that such dynamical collapse models
may indeed be associated with the Class 0 and Class I objects.

As collapse proceeds, infalling material will finally create a centrifugal 
disk whose radius increases in time due to the larger specific 
angular momentum of collapsing material at later times. 
Previous analysis of the growth of the centrifugal disk
(see \cite{cas81,cas83,ter84}) may be modified in the
future to account for the differences between magnetized and unmagnetized 
(exterior) radial collapse models.  Although the theory of the dynamics in
this accretion stage remains rather uncertain, observations seem to 
have identified at least one case  -- the source HL Tau -- where a 
large-scale, radially-infalling ``pseudodisk'' imaged in gas \cite{hay93} 
surrounds a much more compact (presumably centrifugal) disk imaged in dust
\cite{mun96,loo97}.

\subsection*{Accretion in Protostellar Centrifugal Disks}

In the earlier stages of their evolution, protostellar disks are
massive and self-gravity must be important to their evolution.  A
number of workers have investigated analytically and numerically the
growth, development, and saturation of large-scale, low-$m$
gravitational instabilities (i.e. spiral density waves) in disks with
protostellar properties (see e.g.  \cite{lau97} for recent
calculations and references).  Others have taken steps toward
characterizing the action of gravitational torques by an effective
``$\alpha$'' parameter (e.g. \cite{gam97} and references therein);
this approach is more fruitful for local instabilities than global
disk modes.  In addition, still others have discussed how the presence
of binary companions (present in the majority of systems) or planets
can induce accretion through externally-excited gravitational torques
(e.g. \cite{lin93,ost92}).  At present, however, there is no
generalized theory for how the processes of density-wave driven
accretion and binary companion growth may develop in tandem or in
competition out of self-gravitating disk perturbations.  Significant
future efforts will be needed before this stage of disk evolution is
well-understood theoretically.

In the latter stage of centrifugal disk accretion, on the other hand,
disk self-gravity is less important, and we believe we understand the
relevant processes better.  At present, the leading contender for a
physical mechanism to produce the ``viscous'' torques in accretion
disks around compact objects is the saturated Balbus-Hawley instability
(see the contributions of Balbus and Gammie in this volume).  Does the same
process work in the disks of YSOs as well?  The issue here is whether, with the
low temperature and high surface density conditions of protostellar disks, the
gas is sufficiently ionized so that resistive dissipation does not suppress the
instability.  In the inner parts ($R\simlt 0.1$ AU) of protostellar disks,
the central disk temperature exceeds 1000K and collisional ionization of 
alkali metals provides sufficient conduction electrons.  But in the 
low-temperature outer parts of these disks, some source of external ionization
is needed.

Recently, Gammie \cite{gam96} argued that if cosmic rays alone
provide the outer disk ionization, just the upper and lower 
surfaces to a depth of 100 g cm$^{-2}$ will be
sufficiently ionized to couple magnetic fields to the disk matter, and that
the result would be a ``layered model'' where disk surfaces accrete and the
equatorial portion 
is inert.  In corresponding ``alpha'' models of disk accretion, 
the ``active layer'' surface density $\Sigma_a$ is effectively constant with 
$R$, 
while the mass accretion rate $\dot M$ varies with $R$ depending on the local
opacity.  The result is a relatively low value of accretion 
\begin{equation}\label{layer_mdot}
\dot M = 1.8 \times 10^{-8} M_\odot {\rm yr}^{-1} 
\left({\alpha\over 0.01} \right)^2 \left({\Sigma_a\over 100 
{\rm g\ cm^{-3}}} \right)^2 
\end{equation}
in the innermost region, and the possibility of mass accumulation in
the outer disk if the infall rate exceeds the carrying capacity of the
surface layers.  The low predicted value of $\dot M$ in the
``layered'' model appears consistent with recent spectral modeling of 
observed classical T Tauri disks (see the chapter by Calvet).  If the 
difference between the infall rate and accretion rate is large enough, 
the mass buildup in the outer disk could periodically be purged via
gravitational instabilities;  such a mechanism could potentially be
responsible for the observed FU Ori outbursts (see the chapter of 
Calvet for discussion of these phenomena).

Following up on these ideas, Glassgold, Najita, \& Igea \cite{gla97}
have shown that since YSOs are strong X-ray emitters, the surface
layers of their accretion disks will be ionized to a thickness
comparable to Gammie's estimate, even if the cosmic ray flux is
inhibited by strong, magnetized T Tauri winds.  Further work on these
sorts of ``layered'' models will be needed to determine whether both
the observed low mean accretion rate of T Tauri stars and their
high-accretion-rate outbursts events can be explained within a unified
framework.

\section*{Protostellar Winds}

My final topic is the nature winds from
protostellar systems, and their connection to the narrow Herbig-Haro
jets and broader massive molecular outflows which, observationally,
seem to be an inescapable side-effect of protostellar accretion.
Optically-visible Herbig-Haro objects are believed to represent
shocked regions in dense, high-velocity gas that emerges from the very
innermost part of protostellar accretion disks (see contributions by
Frank and Bally in this volume).
Although there is general agreement that the surrounding 
molecular outflows consist of ISM
material that has been mobilized by an interaction with a wind, there is 
equally general disagreement about the nature of that primary wind, and 
its relationship to observed H-H jets.  

In most current models, the primary wind is believed to be
magnetocentrifugally driven, from the inner disk edge where the
accretion flow interacts with the stellar magnetosphere (see \cite{shu94} 
and the review of Stone in this volume),
and/or from the surface of the disk at larger radii (see the
review \cite{kon93}).  The controversial issue is whether the
primary wind consists solely of an isolated, highly collimated flow --
seen as the ionized, optical jet -- or whether there is also a
lower-density, wider-angle, neutral wind surrounding the observed jet.

In the first ``isolated jet'' picture, molecular outflows 
are essentially the wings of bow shocks in the jet head or beam 
(see e.g. \cite{che94});  hence the gas motions transverse to 
the jet are driven by pressure gradients.  One difficulty with such 
models is that the cooling under interstellar conditions may be too strong 
to allow a very broad bow shock to form (e.g. \cite{blo90,sto93}).  
Another potential difficulty is that, depending on the variation of 
magnetic field strength with radius at the base of the flow, it may not even
be possible to drive a wind which is both strongly collimated and fast 
\cite{ost97b}.  A related difficulty is that since 
magnetocentrifugal wind models typically have internal Alfv\'en speeds 
which approach (or even exceed) their flow speeds,
they may not be able to remain collimated as they propagate into the 
low-pressure ambient medium -- where the sound speed is 1-3 orders of 
magnitude smaller than the wind speed.  Adam Frank's contribution 
to this volume describes ongoing numerical studies of jet/cloud interactions
which address these and other issues.

In the second, ``wide-angle wind'' picture, molecular outflows are
interpreted as the pileup from a momentum-driven
``snowplow'' \cite{shu91}.  
A wide-angle wind would occur if a magnetocentrifugal outflow is
unable to self-collimate, or becomes decollimated as it emerges from a
dense core into a medium of low ambient pressure.  Such a wind would
expand laterally to fill the whole solid angle between the
equator and whatever natural boundary exists near the pole.  Because
cold magnetocentrifugal outflows only occur on field lines having
angles less than $60^\circ$ with respect to the equator, it is natural
to expect that there would remain some unloaded axial fields filling
the space interior to the wind towards the pole, which originate in
the disk at angles greater than $60^\circ$ with respect to the equator.
In particular, when the wind is formed by tearing open a
magnetosphere \cite{shu94,ost95}, the axial magnetic flux interior to the
wind will be comparable to the poloidal magnetic flux in the wind itself.
If the wind field lines are instead open field lines carried in with the
accretion flow, then the ratio of interior magnetic flux to wind magnetic 
flux would depend on the distribution of the field in the disk.

The asymptotic structure of a wide-angle MHD wind, for the case of a
flow along opened magnetospheric field lines emerging nearly
isotropically from the corotation region (the ``x-wind''), has been
analyzed by Shu, Najita, Ostriker, \& Shang \cite{shu95b}.  
They show that, while the streamlines
collimate only slowly with distance, the density structure in the wind
becomes nearly cylindrically stratified.  Briefly, the reasons for these
results can be explained as follows: In the asymptotic regime, the
largest term in the internal stress is that associated with the
toroidal magnetic field $B_\varphi$.  In order to have a locally
force-free configuration, $\partial (B_\varphi R)/\partial \theta
\approx 0$ must hold, where $R=r\sin\theta$ is the cylindrical
distance from the axis, and $\theta$ is the polar angle.  This implies
\begin{equation}\label{force_eq}
B_\varphi =-C(r)/R
\end{equation}
for some (positive) function $C(r)$.
From angular momentum conservation
and field freezing, we have 
\begin{equation}\label{angmom_eq}
{B_\varphi\over B_p} = -{\Omega R\over v_p} \hskip1cm {\rm and}\hskip1cm B_p=\beta\rho v_p,
\end{equation}
where $v_p$ and $B_p$ are respectively the poloidal flow speed and
poloidal magnetic field, $\Omega$ is the angular rotation rate of the
footpoint of the local field line, $\rho$ is the density, and $\beta$
represents the ratio of magnetic flux to mass flux (which remains
constant on any streamline).  From these equations, 
we find
\begin{equation}\label{rhosol}
\rho={C(r)\over \Omega\beta R^2}.
\end{equation}
Using ${\bf \nabla\cdot B_p}=0$, it can be shown that $C$ is a slowly-varying
function of $r$.  Thus, if $\beta$ and $\Omega$ vary little between
streamlines (as is true in the x-wind solutions of Najita and Shu 
\cite{naj94}), then
the density in the wind will vary approximately as $\rho\propto
R^{-2}$ at large distances.  Note that while the scaling $\rho \propto r^{-2}$
arises kinematically
from the continuity equation for a constant-speed
radial flow, the scaling $\rho\propto (\sin\theta)^{-2}$ comes about because 
hoop stresses redistribute the streamlines in $\theta$ until 
the  magnetic field is force-free (cf. eq. \ref{force_eq}).
The actual value of the function $C(r)$ may be evaluated by
balancing the pressure at the inner edge of the wind
$B_\varphi^2/(8\pi)$ against the pressure provided by interior axial
fields, $B_{p, int}^2/(8\pi)$, and applying the constraint of mass
outflow conservation.  The value of the poloidal speed on any
streamline can be obtained from energy conservation; in terms of the
streamline's footpoint radius $R_0$ and Alfv\'en radius $R_A$,
$v_p=\Omega (2 R_A^2 - 3 R_0^2)^{1/2}$.

For any variation of $\Omega$, $\beta$, $R_0$, and $R_A$ with magnetic
flux $\Phi$ (this variation must be derived self-consistently for the
wind inside the fast-MHD surface), the foregoing considerations can be
applied to find the asymptotic structure of a ``free'' MHD wind --
whether the wind originates in a narrow or broad region of the disk.
It can be shown, for example, that disk winds which leave the disk
with $\rho\propto R_0^{-q}$ and $B\propto R_0^{-(1+q)/2}$ also have
slowly-collimating (i.e. pointing close to radially) streamlines but
nearly cylindrically-stratified density structure near the pole, at
large distances from the disk.  Figure 1 shows an example of the
asymptotic structure of a wind in which all streamlines originate at
the same radius $R_0$ and have $R_A=2 R_0$, and for which the interior
axial magnetic flux along the pole equals the wind poloidal magnetic
flux.

\begin{figure}[hb] 
\centering
\epsfig{file=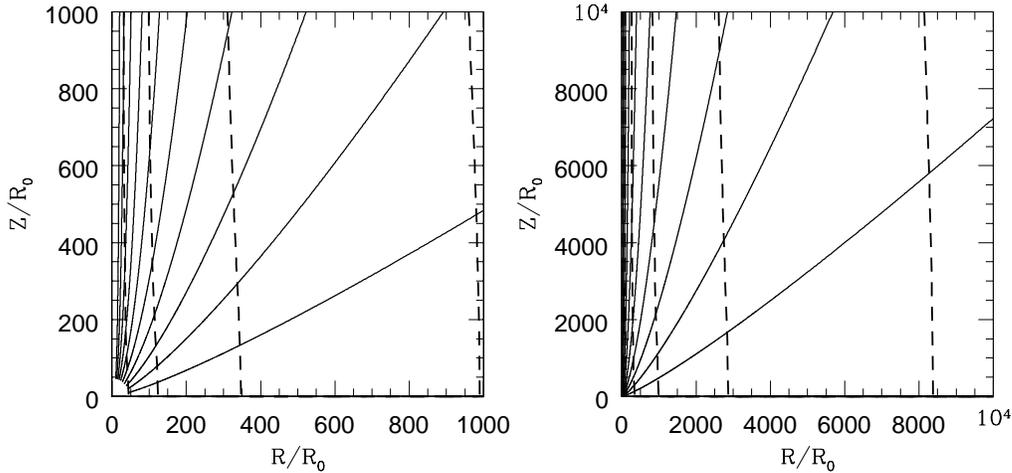,width=5.5in}
\vspace{-2.5in}
\caption{
Asymptotic magnetocentrifugal wind, on two scales.  Streamlines/field lines 
are shown with solid curves at equal increments of mass loss/magnetic flux.  
Density contours are shown with dashed curves, for $log(\rho/\rho_0)
= -4, -5, -6, -7$  (left), and 
$log(\rho/\rho_0)= -6, -7, -8, -9$  (right); 
$\rho_0\equiv\dot M_w/(4\pi R_0^3 \Omega)$.  
}
\end{figure}

Wide-angle ``snowplow'' models with winds of this sort 
have been successful at explaining many
general characteristics of outflows \cite{li96}, as well as particular outflow 
systems \cite{nag97}.  

Further work on both observational and theoretical fronts is still needed to
identify which set of conditions leads to a wide-angle wind, and which to
an isolated jet, or perhaps to conclude instead that alternative models are
warranted to represent the primary winds from protostars.

I am grateful to Charles Gammie and Jim Stone for their helpful comments on
this manuscript.

\end{document}